\documentclass[12pt]{iopart}
\usepackage{amsmath}
\usepackage[dvips]{graphicx}
\usepackage{dcolumn}

\newcommand{\lbrp}{\left| }
\newcommand{\rbrp}{\right| }
\newcommand{\lbrla}{\left\langle }
\newcommand{\rbrra}{\right\rangle }
\newcommand{\half}{\frac{1}{2}}
\newcommand{\bra}[1]{\lbrla #1 \rbrp}
\newcommand{\ket}[1]{\lbrp #1 \rbrra}
\newcommand{\braket}[2]{\lbrla #1 | #2 \rbrra}
\newcommand{\D}{\text{d}}
\begin{document}

\title[Band selection and disentanglement]
{Band selection and disentanglement using maximally-localized Wannier
functions: the cases of Co impurities in bulk copper and  the Cu (111) surface}

\author{Richard Koryt\'ar$^1$}
\ead{rkorytar@cin2.es}
\author{Miguel Pruneda$^1$}
\author{Javier Junquera$^2$}
\author{Pablo Ordej\'on$^1$}
\author{Nicol\'as Lorente$^1$}
\address{$^1$Centre d'Investigaci\'o en Nanoci\`encia i Nanotecnologia
 (CSIC - ICN), Campus de la UAB,
     E-08193 Bellaterra, Spain}
\address{$^2$Departamento de Ciencias de la Tierra y F\'{\i}sica
de la Materia Condensada,
Universidad de Cantabria,
     E-39005 Santander, Spain}
\date{\today}
\begin{abstract}
We have adapted the maximally-localized Wannier function approach of
[Souza I, Marzari N and Vanderbilt D 2002 \emph{Phys. Rev. B} {\bf 65} 035109]
to the density functional theory based {\sc Siesta}  code
[Soler J M {\em et al.} 2002 \emph{J. Phys.: Cond. Mat.} {\bf 14} 2745]
and applied it to the study of Co substitutional impurities in bulk 
copper
as well as to the Cu (111) surface. In the Co impurity case, we have
reduced the problem to the Co $d$-electrons and the Cu $sp$-band, permitting
us to obtain an Anderson-like Hamiltonian from well defined density functional
parameters in a fully orthonormal basis set. In order to test the quality of 
the Wannier approach to surfaces, we have studied the electronic structure of
the Cu (111) surface by again transforming the density functional problem into
the Wannier representation. An excellent description of the Shockley surface state is
attained, permitting us to be confident in the application of this method to
future studies of magnetic adsorbates in the presence of an extended surface state.
\end{abstract}
\maketitle

\section{Introduction}

The advent of density functional theory (DFT) in electronic
structure calculations has revolutionized the fields of quantum
chemistry and, more generally, of condensed matter physics~\cite{Martin}. However,
due to necessary approximations of the unknown functionals
(typically, the local density approximation or LDA, and semilocal approximations as the generalized gradient
approximation or GGA),
important limitations prevail in the description of many systems
ranging from insulating materials to impurities in a metal host. The
general approach to these problems is to go beyond DFT
by proposing a simplified
Hamiltonian that can be solved, at the expense of losing the parameter-free
advantage of DFT. Recently, the situation is changing and new methods
either stemming from DFT or making use of DFT for initial input are
emerging~\cite{Kotliar,Katsnelson}. 
One such example is the LDA+U approach~\cite{Anisimov91,Solovyev}
where the exchange-correlation potential for the local electron
gas is complemented by missing strong localized correlations. The
success of this approach has been considerable in explaining
the opening of a band gap in systems with localized electronic states. Yet,
the choice of the Coulomb intra-atomic energy $U$ is
somewhat arbitrary, depending on the choice of basis sets or
other descriptors of the treated system~\cite{Cococcioni}.

Nakamura and co-workers~\cite{Nakamura}
have developed a constrained LDA approach based on
maximally localized Wannier functions (MLWF)~\cite{PRB_Mlwf_Souza}.
The MLWF replace the linear muffin-tin
orbitals (LMTO) that were initially used as a natural
basis set to define $U$~\cite{Anisimov91,Gunnarsson89,Gunnarsson91}.
As Nakamura {\em et al.} emphasize~\cite{Nakamura}, the properties
of MLWF are particularly appealing for reducing the complicated
DFT problem to a simplified Hamiltonian, where especial physics
can be explored such as the localized electron correlations mentioned above.
Wannier functions have been thoroughly explored in the
work by Marzari and Vanderbilt~\cite{Marzari}, and an algorithmic
approach to obtaining them from a DFT calculation is available. 
More recently, Souza, Marzari and Vanderbilt~\cite{PRB_Mlwf_Souza}
have extended the approach to disentangle electronic bands,
creating a compact local basis set that accurately
reproduces the DFT electronic structure in a given energy
window. This approach
is independent of the actual implementation of the DFT calculation,
yielding a natural way of describing an extended basis set calculation
in terms of localized functions. 

In the present work, we have interfaced the approach by Souza 
{\em et al.}~\cite{PRB_Mlwf_Souza} as implemented in {\sc Wannier90}~\cite{CPC_Wannier90}
 to the {\sc Siesta} code~\cite{Soler}.
Contrary to the case of Nakamura {\em et al.}~\cite{Nakamura}, {\sc Siesta} is
an atomic basis set code, and it would seem natural to use
the atomic orbitals  to define $U$ and  associated model Hamiltonians.
However, besides their optimized spread, MLWF have two important
features: $(i)$ they are a naturally orthogonal basis set,
rendering tight-binding like approaches easy to use $(ii)$ the
extraordinary accuracy of MLWF in a defined energy window permits to
disentangle electronic bands~\cite{PRB_Mlwf_Souza} and to have
a simple tight-binding approach with DFT accuracy within the chosen
energy window. Hence, our present implementation of MLWF
permits us to translate the  complex atomic orbital  method into a simple
orthogonal tight-binding approach that can be easily cast into
an Anderson-Hamiltonian model~\cite{Anderson} or a Hubbard one~\cite{Hubbard}.

The manuscript is organized in four sections. The methods section deals
with a small overview of the work of reference~\cite{PRB_Mlwf_Souza}, and
the technical details of the actual  numerical  implementation. The
approach is then applied to two different systems: $(i)$
a substitutional Co impurity
in bulk Cu and $(ii)$ the Cu (111) surface. Both systems
are of uttermost interest. The case of Co in Cu is a classical
example where strong correlations are in play
leading to very large Kondo temperatures~\cite{Lang,Quaas}, hence it is
interesting to understand the magnetic properties leading to the Kondo
correlations in this system~\cite{Lichtenstein}. Besides, Co in Cu is one of the model
giant magnetoresistance (GMR) systems~\cite{Dederichs,Zahn,Fan}.
The concentration of Co in Cu is an important parameter in its magnetic
properties~\cite{Fan}. Indeed, when the Co to Cu atomic ratio is above
1 to 4, the system becomes a ferromagnet. Below this density, 
Fan {\em et al.}~\cite{Fan} experimentally find
that the Co atoms are distant enough to show paramagnetism in agreement
with the paradigmatic analysis by Goodenough~\cite{Goodenough}. We will concentrate
in this paramagnetic phase, with two different densities, for  $2\times2\times2$
and  $4\times4\times4$ cubic supercells.

Our calculations
are a first step in the study of the complex electronic structure of
the Co-impurity  problem that presents a high ($\sim 500$ K)
Kondo temperature~\cite{Quaas}.
Hence, the evolution of the electronic structure as the Co concentration
is reduced, permits us to study the effect of Co concentration and the
use of MWLF gives us access to an Anderson-like Hamiltonian.

The study of the Cu(111) surface is also very interesting given the
existence of the $LL'$ gap ($\bar\Gamma$ in the surface Brillouin zone)
and the associated Shockley state. Despite the 
locality of the MLWF, the calculations succeed in accounting
for the  surface Shockley state and in yielding  the correct
electronic structure about the Fermi energy. 

This work shows that
MLWF are accurate enough to study the magnetic properties of
Co on Cu(111) and are a first step towards the study
of the electronic structure~\cite{Bluegel} on surface systems as well as
the Kondo effect of magnetic adsorbates that has recently
received much experimental~\cite{Knorr,Vitali} and theoretical~\cite{Carter}
attention.

\section{Method}

\subsection{Wannier functions and {\sc Wannier90}}

Wannier functions, $w(\mathbf{r-R})$, are formally identical to
Fourier coefficients of Bloch waves,
$\psi_{\mathbf k}(\mathbf r)$, given by
\begin{equation}
w(\mathbf{r-R}) = \frac{1}{\Omega^\star}\int_{BZ}
\psi_{\mathbf k}(\mathbf r) e^{-i\mathbf{k \cdot R}}\, \D^3 \mathbf{k}.
\end{equation}
Here $\mathbf R$ denote Bravais vectors and $\Omega^\star$ is the volume
of the Brillouine zone (BZ). This definition suffers from the indeterminacy
of the phases of Bloch functions, $\psi_{\mathbf k}(\mathbf r)$.
Furthermore, if a group of bands, $\{n\}$ is considered, with corresponding Bloch
functions, $\psi_{\mathbf kn}(\mathbf r)$,
it is ambiguous to talk about an individual band and the above definition
can be generalized to
\begin{equation}
w_m(\mathbf{r-R}) = \frac{1}{\Omega^\star}\int
\sum_n U_{mn}(\mathbf k)
\psi_{\mathbf kn}(\mathbf r) e^{-i\mathbf{k\cdot R}}\, \D^3\mathbf{k},
\end{equation}
where $U_{mn}(\mathbf k)$ is an arbitrary unitary matrix. The unitarity
guarantees that the Wannier functions will be orthogonal. 
The arbitrariness of  $U_{mn}(\mathbf k)$ allows for tuning
the phases of Bloch functions in the integral as well as the admixture of functions
pertaining to different bands.
Thus, there is a whole class of Wannier functions for the given band structure.
Marzari and Vanderbilt devised a variational scheme that determines the
Wannier functions with minimum total spread, $\Omega$ defined as
\begin{equation}
\Omega = \sum_m\left[\langle r^2\rangle_m - \langle r\rangle_m^2\right],
\label{omega}
\end{equation}
where we have used the notation
$\langle\ \cdot\ \rangle_m = \bra{w_m}\ \cdot\ \ket{w_m}$.
For a given set of Bloch functions,
the total spread, $\Omega$, is a functional of
the unitary matrices $U_{mn}(\mathbf k)$. The Wannier functions so obtained are
called maximally-localized Wannier functions (MLWF). 

The locality and the orthogonality
of Wannier functions permit us to have a straightforward compact tight-binding representation
of the Hamiltonian in systems where the group of bands of interest
is separated from the rest of the electronic structure 
by a band gap. In metals, the absence of band gaps renders
the separation of states more complicated. Souza {\em et al.} have
devised a disentanglement procedure~\cite{PRB_Mlwf_Souza}
by focusing on a certain energy window, hereinafter called {\em outer} energy window, 
and by selecting certain bands in it.
Hence, the disentanglement procedure necessarily reduces the number of
states inside the {\em outer} energy window, while exactly reproducing
the electronic bands in a certain energy
interval called {\em inner} energy window. 
The selection of bands proceeds via 
trial orbitals $g_n(\mathbf r)$ 
that allow to define the character of
the Wannier subspace of interest. 

The numerical calculations of MLWF have been done with the 
{\sc Wannier90}~\cite{CPC_Wannier90} code.
This package can be used as a post-processing tool
with most first-principles codes.  In practical simulations, the  Bloch
states,
$ | \psi_{\mathbf{k} n} \rangle $,
are computed in a mesh of uniformly spaced $\mathbf{k}$-points
within the first-Brillouin zone.  The basis set used to represent the 
wave functions changes from one electronic structure code to another one; however,
the MLWF algorithm 
requires an input that is essentially basis-independent. In particular, the main
ingredients are (i) the overlap matrix between the cell-periodic parts
of the wavefunctions at neighboring $\mathbf{k}$-points
 \begin{equation}
\label{Eq_M_matrix}
  M_{mn} (\mathbf{k}, \mathbf{b}) = \langle u_{\mathbf{k} m} | u_{\mathbf{k} + \mathbf{b} n}\rangle =
     \langle \psi_{\mathbf{k} m} | e^{-i \mathbf{b} \cdot \mathbf{r}}|
             \psi_{\mathbf{k} + \mathbf{b} n}\rangle,
\end{equation}
\noindent (ii) the Bloch energies $\varepsilon_{\mathbf{k} n}$
on the regular grid of $\mathbf{k}$-points and
 \noindent (iii) the coefficients of the above trial orbitals, $g_{n}(\mathbf r)$, in the  Bloch basis
\begin{equation}
\label{Eq_A_matrix}
     A_{mn} (\mathbf{k}) = \langle \psi_{\mathbf{k}m} | g_{n}\rangle.
\end{equation}
The trial functions we employ are of the form 
$g_n(\mathbf r) = R_n(r)\Theta_{lm_r}(\phi,\theta),\ \mathbf r=r,\phi,\theta$,
a product of nodeless hydrogenic-like radial part $R_n(r)$
and a real spherical harmonic $\Theta_{lm_r}(\phi,\theta)$ with angular momentum $l$
and projection $m_r$. 

From this, the {\sc Wannier90} code computes the final unitary
transformation matrices $U_{mn} (\mathbf{k})$.  
With the aid of the matrices $U_{mn}(\mathbf k)$ the
Hamiltonian can be expressed in the Wannier basis and diagonalized at any
k-point.  These Wannier-based energy bands are
further compared to the initial  \emph{ab-initio} bands in order to test
the quality of the newly obtained Wannier basis set.

It is very useful to work with
density of states projected onto a certain Wannier function of spin $\sigma$,
$w_{m\sigma}$. This projected density of states (PDOS)\cite{Lorente} is
the spectral function, $\rho_{m\sigma}(\omega)$, given by
\begin{align}
\rho_{m\sigma}(\omega) &= \sum_{n\mathbf k}
\left|\braket{w_{m\sigma}}{\psi_{\mathbf kn\sigma}}\right|^2
\delta(\omega - \epsilon_{\mathbf kn\sigma}) \nonumber \\
&=\sum_{n\mathbf k}\left|U_{mn}(\mathbf k\sigma)\right|^2
\delta(\omega - \epsilon_{\mathbf kn\sigma}).
\label{PDOS}
\end{align}
This 
Wannier PDOS or spectral function  shows how the Wannier character
is distributed in energy over the electronic band structure~\footnote{The
PDOS can be used to analyze any electronic structure by choosing the analyzing 
functions, for example, molecular orbitals were used to describe the electronic structure of 
benzene on Cu(100) in \cite{Lorente}.}.

From \eref{PDOS} it is straightforward to define the Wannier function
occupation  at zero temperature as
\begin{equation}
\label{Eq_def_occ}
n_{m\sigma} = \int_{-\infty}^\mu\rho_{m\sigma}(\omega)\D\omega,
\end{equation}
where $\mu$ is the Fermi level.

\subsection{Pseudo-atomic orbital DFT calculations}

\emph{Ab-initio} DFT calculation presented in this work
are based on strictly localized~\cite{PRB_Sankey} numerical pseudo-atomic
orbitals (PAO)~\cite{PRB_Junquera} that are solutions to the atomic Kohn 
and Sham equation with norm-conserving pseudopotentials.
In particular, the calculations are done using the
\textsc{Siesta}~\cite{Soler} package.

The matrix elements between basis functions are calculated by 
real-space integration, and the
Hamiltonian eigenstates are labeled using the Bloch theorem,
because periodic boundary conditions are imposed. The Bloch functions can
be expanded into the PAO basis as follows
\begin{equation}
\label{Eq_Bloch_expansion}
\psi_{n\mathbf k}(\mathbf r) = \sum_{\mathbf R\mu}c_{\mu n}(\mathbf k)
e^{i\mathbf{k\cdot(r_\mu+R)}}\varphi_\mu(\mathbf{r-r_\mu-R})
\end{equation}
where we assume that the unit cell contains the centers $\mathbf r_\mu$
of basis functions $\varphi_\mu(\mathbf{r-r_\mu})$ which are then
repeated periodically to every other cell by the Bravais lattice
vector $\mathbf R$. Finally, the complex numbers $c_{\mu n}(\mathbf k)$
are expansion coefficients.
The Brillouin zone is sampled uniformly.
For the exchange and correlation potential entering
the Kohn and Sham equation, 
we use the PBE generalized gradient approximation 
\cite{PRL_PBE}.
The chosen atomic basis set is an optimized double-$\zeta$ plus polarization
  for the valence states of Co and Cu, amounting to 15 basis functions per atom.  All the parameters that
  define the shape
  and the range of the basis functions were obtained by a variational
  optimization of the enthalpy (energy plus a penalty for orbital 
 volume increase) with a pressure of $P = 0.1$ GPa,
  following the recipe given in reference~\cite{PRB_Anglada}.
The substitutional Co impurity basis set was optimized in the Cu host,
and the Cu basis set corresponds to the optimal one for bulk Cu.
The question of optimal basis sets for surfaces is more intricate and
has been discussed in \cite{PRB_Sandra}. We use the 
basis set obtained in reference~\cite{PRB_Sandra},
which is based on the same enthalpy minimization
with an especial focus on the vacuum extension of
the surface state density.

\subsection{Implementation of Wannier functions}

The implementation of maximally localized Wannier functions in \textsc{Siesta}
consists in evaluation of (\ref{Eq_M_matrix}) and (\ref{Eq_A_matrix}).
Expanding the Bloch functions according to 
\eref{Eq_Bloch_expansion}, we obtain

\begin{equation}
\label{Eq_Mkb}
M_{mn}(\mathbf k,\mathbf b) = 
\sum_{\mu\nu}\sum_\mathbf{R}c^*_{\mu m}(\mathbf k)c_{\nu n}(\mathbf{k+b})
e^{i\mathbf{k\cdot (R - r_\mu + r_\nu)}} \mathcal M_{\mu\nu}(\mathbf R,\mathbf b)
\end{equation}
along with
\begin{equation}
\label{Eq_Ak}
A_{mn}(\mathbf k) = \sum_\mu c_{\mu m}^\ast(\mathbf k)\sum_{\mathbf R}
e^{-i\mathbf k\cdot\left(\mathbf R+\mathbf r_\mu\right)}\mathcal A_{\mu n}(\mathbf R),
\end{equation}
where
\begin{equation}
\label{Eq_m}
\mathcal M_{\mu\nu}(\mathbf R,\mathbf b) =
\int \varphi_\mu^*(\mathbf{r+R - r_\mu + r_\nu})e^{-i\mathbf{b\cdot r}}
\varphi_\nu(\mathbf r)\D^3\mathbf r
\end{equation}
and
\begin{equation}
\label{Eq_a}
\mathcal A_{\mu n}(\mathbf R) =
\int\varphi^\ast_\mu(\mathbf{r-r_\mu-R})g_n(\mathbf r)\D^3\mathbf r.
\end{equation}

This reduces the computation of (\ref{Eq_M_matrix}) and (\ref{Eq_A_matrix}) to the calculation of a few matrix elements of 
localized functions, equations~(\ref{Eq_m}) and (\ref{Eq_a}).
The first integral is computed on the real space grid
while the second integral makes use of the analytic angular dependence of 
the integrand in the same way as is done for the calculation of 
overlap matrices in reference\cite{Soler}.
This is an important difference with the implementation of ~\cite{Weng}
where an expansion on powers of the integrand is performed. Another difference
of our implementation is that we write an interface for \textsc{Wannier90}, and hence
the trial functions are the ones for \textsc{Wannier90}, while in 
reference~\cite{Weng}
the original basis set is used.

Finally, we note that Brillouin zone sampling used to obtain the 
self-consistent
electronic density in \textsc{Siesta} is essentially independent 
from the sampling
used for the input data of \textsc{Wannier90},  {\em i.e.}
Equations~(\ref{Eq_M_matrix},\ref{Eq_A_matrix}) and $\epsilon_{n\mathbf k}$.

\subsection{\label{Sec:2D}
\emph{Ab-initio} calculation of a cobalt impurity in bulk copper}

The first system, the cobalt impurity in a bulk FCC copper host matrix,
was simulated by a supercell where one host atom was replaced by a cobalt
one. The lattice
parameter was fixed to the theoretical 
value we found for pure copper ($a_{lat}=3.690$\AA).
We use two different unit cells to represent the system: $(i)$ an eight-atom
($2\times2\times2$) cell~\footnote{
For the supercell of 8 atoms ($2\times2\times2$), the {\sc Siesta} calculation parameters are the
following:  the Brillouin zone sampling was $5\times5\times5$, 
the discretization of the real-space mesh was taken from the mesh cutoff value 500 Ry and 
the self-consistency tolerance of the density matrix was $10^{-4}$.
}
, where there is a $1/7$ ratio of Co impurities,
and $(ii)$ a 64-atom one ($4\times4\times4$)~\footnote{
For the supercell of 64 atoms ($4\times4\times4$), the Brillouin zone sampling was $3\times3\times3$, 
keeping the same convergence criteria as for the $2\times2\times2$-cell. }
where the cobalt concentration
drops to $1/63$. Besides its lower computational demand,
the small cell permits us to represent the calculated
bands and follow the Wannier disentanglement in a simpler way
avoiding the larger band folding of the ($4\times4\times4$) supercell. 
 According to the experimental data by Fan {\em et al.}~\cite{Fan}
only above  a concentration ratio of $1/4$ 
does Co in Cu show  ferromagnetic ordering; below this concentration, 
the system becomes paramagnetic. 
Our calculations correspond to impurity densities of the paramagnetic phase. 

Interestingly, the measured saturation magnetic moment per atom is 
1.5 $\mu_B$ for  pure Co
and it drops to 0.4$\mu_B$/atom for a $1/9$ concentration~\cite{Fan}.
Our calculations yield 0.15$\mu_B$/atom for the $1/7$ concentration, indicating a smaller
spin-polarization of our DFT calculations. It is difficult to know the actual experimental
error bar, but our calculations yield the right order of magnitude as well as the trend
with decreasing Co concentration. Indeed, the supercell magnetization for the $2\times2\times2$-case
($1/7$ concentration) is 1.03 $\mu_B$ (0.15$\mu_B$/atom)
 while the supercell magnetization
for the $4\times4\times4$ is 1.57 $\mu_B$ (0.025$\mu_B$/atom), indicating a saturation of the cell magnetization
and thus a decrease of magnetization per atom, roughly linear with the number of electrons
following the Slater-Pauling curve~\cite{Fan}.  

For the obtention of the MLWF, 55 bands in the $2\times2\times2$ supercell
 and 460 in the case of a $4\times4\times4$ supercell,
were extracted for both spins,
spanning the energy interval of (-14.0,9.1) eV 
with respect to  the Fermi level
(fully including the lowest valence bands). This energy interval 
is the {\em outer} window of \textsc{Wannier90}.

At the Fermi surface, the relevant states are both dispersive conduction bands
and hybridized cobalt states originating from the Co incomplete $d$-shell.
On the other hand, the closed copper $d$-shell gives rise to narrow bands, lying
deeper below the Fermi energy ($\mu$). In order to simplify the host
electronic structure,  we want these Cu d-bands to be projected out during the
disentanglement process
and  keep a simpler $sp$-like band. This can be achieved by 
both
(1) the choice of the {\em inner} window position that
freezes the bands with dispersive character around $\mu$ and 
(2) the  choice of trial
orbitals. As for the latter, we used one spherically symmetric
function per atom, located 
in one of the two interstitials of the fcc primitive cell. In addition, five
cobalt centered orbitals with angular momentum $l=2$ were used.
The {\em inner} window choice is discussed in Section~\ref{Sec_Results}.
Finally, for the spread, $\Omega$, minimization a tolerance of 
$10^{-10}$ \AA$^2$~ has been used.

\subsection{Simulations of Cu (111)}

The second system we studied is the Cu (111) surface. We used a  12-atom
slab with a $1 \times 1$ surface unit cell. The k-point sampling
 was $14\times14\times1$ and 
real space mesh cutoff of 800 Ry was used. Please, refer
to \cite{PRB_Sandra} for more details.
The Bloch functions corresponding to the lowest 120 bands were used 
to calculate
the matrix $M_{mn}(\mathbf{k,b})$ using a k-point sampling $10\times10\times1$
that largely suffices for convergence obtaining the MLWF.
As in the previous case,
we are interested in reproducing with the simplified scheme of MLWF the electronic
structure about the Fermi energy, $\mu$, hence we limit the {\em inner} energy window 
to the energy interval $\langle -1.0,2.0\rangle $ eV.
The {\em inner} energy window must not contain more bands than the 
required number of
Wannier functions, so the
lower limit of the {\em inner} energy window is chosen 
somewhat above the narrow d-like bands and its upper
edge of the {\em inner} window is limited by the 
higher-energy dispersive bands.

\begin{figure}
\includegraphics[scale=0.43]{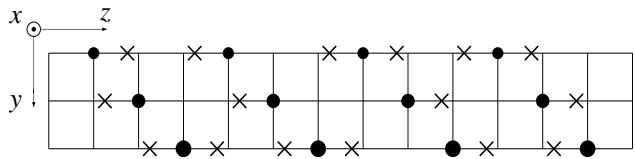}

\caption{\label{Fig_trial_scheme}
Scheme of the unit cell of the slab.  Atomic basis functions
are centered on Cu atoms (dots). Hence, the dots reproduce
the slab geometry, where the size of the dots conveys information
on the
different x-coordinate. Crosses indicate the center for
further \textsc{Wannier90} orbitals, the
 \textsc{Wannier90}
spherical interstitial trial orbitals, each located in the
$\frac{1}{4}$ and $\frac{3}{4}$ of the bulk (111) translation vector,
which points along the z-axis of the scheme.}

\end{figure}

Among various configurations of trial orbitals, we found that the one described in
figure~\ref{Fig_trial_scheme} converges to good-quality MLWF. 
The figure
shows the centers of the Wannier functions that are located
both at the atomic positions of the slab (dots) and in between
(crosses) to enhance the electronic structure description.
Inside the slab, one spherical
interstitial is sufficient to give a good description of the electronic structure
close to the Fermi energy, $\mu$. 
On the surface, two functions are necessary; we put them  at the
positions corresponding to the interstitials of the bulk geometry.
The trial orbital set has been found stable: only the outermost Wannier
functions move some $0.21$~\AA~outwards during the spread minimization.

\section{Results and discussion}
\label{Sec_Results}

\subsection{Cobalt impurity embedded in bulk copper}

In this section, we present and discuss MLWF for the impurity system.
The essence of what we are doing is a tight-binding representation of a
conduction band (supported by one interstitial Wannier function per atom)
hybridizing with impurity Wannier functions. This picture will be extended in the
next section, where we attempt to separate the Coulomb interaction in the impurity
in the spirit of Anderson Hamiltonian.

\begin{table}
\caption{\label{Tab_energies_2}
Spread ($\Omega$), on-site energies ($\widetilde\epsilon$) and occupation ($n$)
for the cobalt Wannier functions of a substitutional Co atom in the $2\times2\times2$-Cu cell.
These split into two- and three-fold degenerate states $m=\{e_g,t_{2g}\}$.
Majority ($\sigma = \uparrow$) and
minority ($\sigma = \downarrow$) spins are given.}
\begin{indented}
\item[]
\begin{tabular}{@{}llllll}
\br
 & &  $e_g^\uparrow$  &  $t_{2g}^\uparrow$  &  $e_g^\downarrow$  &  $t_{2g}^\downarrow$  \\
\mr
 $\Omega_{m\sigma}$       & [\AA$^2$]      & 1.22  & 3.60    & 1.66  & 2.95\\
 $\widetilde\epsilon_{m\sigma}-\mu$ & [eV] &-1.80  &-0.74    & 0.24  &-0.01 \\
 $n_{m\sigma}$            & [e]            & 0.94  & 0.82    & 0.36  & 0.74 \\
\br
\end{tabular}
\end{indented}

\end{table}

\begin{table}
\caption{\label{Tab_energies_4} Spread ($\Omega$), on-site energies ($\widetilde\epsilon$) and occupation ($n$)
for the cobalt Wannier functions of a substitutional Co atom  in a $4\times4\times4$-Cu cell.
As in the previous case, the symmetry partially
removes the degeneracy and now, two types of states
are found, 
doubly and threefold degenerate:  $m=\{e_g,t_{2g}\}$ for the majority ($\sigma = \uparrow$) and
minority ($\sigma =  \downarrow$) spins.}
\begin{indented}
\item[]
\begin{tabular}{@{}llllll}
\br
 && $e_g^\uparrow$ & $t_{2g}^\uparrow$ & $e_g^\downarrow$ & $t_{2g}^\downarrow$ \\
\mr
  $\Omega_{m\sigma}$       & [\AA$^2$] &  0.82 &  1.28  & 1.80  &  3.11   \\
  $\widetilde\epsilon_{m\sigma}-\mu$ & [eV]    & -2.16 & -1.45  & 0.23  & -0.02   \\
  $n_{m\sigma}$            & [e]       &  0.96 &  0.89  & 0.33  &  0.67   \\
\br
\end{tabular}
\end{indented}

\end{table}

\begin{figure}
\includegraphics[scale=0.7]{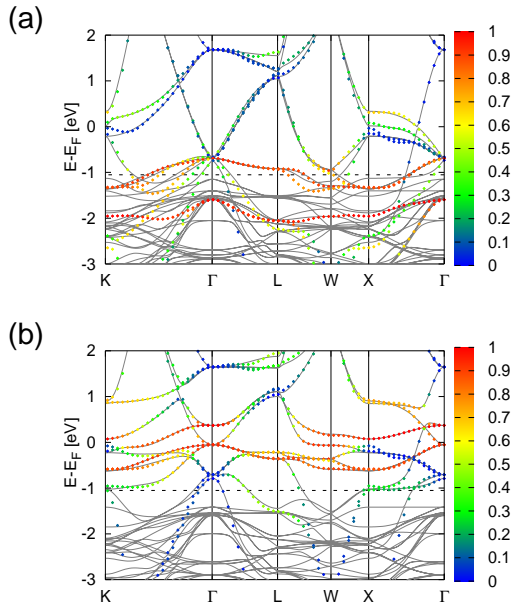}
\caption{\label{Fig_spd_bands_2}(Color online) Interpolated band structure for bulk copper 
 with a substitutional cobalt impurity in an eight-atom cell ($2\times2\times2$),
(a) majority spin, (b) minority spin.
 Color denotes overlap of the eigenstate with cobalt states.
 For comparison, the {\em ab-initio} structure is plotted in grey.
 The zero of energy is the Fermi level.
 The {\em inner} energy window is the upper part of each diagram starting at the 
 dashed line at $-1$ eV.}
\end{figure}

Figure~\ref{Fig_spd_bands_2} shows the  \emph{ab-initio}  band structure and the 
disentangled one for the $2\times2\times2$-cell. 
The color code denotes 100\% overlap with a Co MLWF for red, and 0\% for blue.
The chosen
{\em inner} energy window starts at $-1.05$ eV and extends up to $+3.6$ eV.
This, in combination with trial orbital choice, efficiently disentangles
the Cu $sp$-bands from the Cu d-bands, removes these last ones, 
and keeps the rest together with the full $d$-electron structure
of the Co impurity.
As we described above, the spin polarization of the system is sizable due to the
presence of Co atoms.
This is clearly seen in the present graph, where the bands split according to
their spin, 
in figure~\ref{Fig_spd_bands_2} (a), the majority spin is represented, and 
correspond to Co $d$-bands  
that coexist in the region of the Cu $d$-bands. The quality of these bands 
stemming from the MLWF
calculation is considerably worse than the minority spin ones,
figure~\ref{Fig_spd_bands_2} (b), because
they lie outside the {\em inner} energy window. Nevertheless, the calculation 
contains 
information on the effect of the Cu d-bands on the majority spin Co bands
as we will discuss in the analysis of the PDOS.
Moreover, the
calculations dealing with the magnetic structure and other
information close
to the Fermi energy will be very accurate as can be seen 
in the excellent matching
of the  \emph{ab-initio} bands
and the MLWF ones near the Fermi level, figure~\ref{Fig_spd_bands_2}.
The calculation retains the
Cu $sp$-bands that correspond to the colder-colored bands of 
figure~\ref{Fig_spd_bands_2} that strongly
disperse. However, the hot-colored bands present flat features,
revealing a small Co-Co
intercell interaction, revealing that the dispersion largely comes
from near-neighbor interactions, Co-Cu. 
For the $4\times4\times4$ cell this interaction is even smaller, 
but the bands are not much flatter, what implies that
for many Co-based properties our $2\times2\times2$ calculations will suffice.  

It can be seen that the MLWF disentanglement projects out the Cu $d$-bands 
but retains a realistic description of Co $d$-bands, 
as revealed
by looking into the spin polarization of the system.
Indeed, by evaluating (\ref{Eq_def_occ}), we
can compute the electronic population of the different MLWF.
For the $4\times4\times4$ cell, the majority spin MLWF
yields 4.59 electrons, while the minority spin gives a population of 2.65r;, if we
add the spin polarization of the remaining Cu bands, we obtain 1.72 electrons,
 in  good
agreement with the 1.57 electrons of the full \emph{ab-initio} calculation,
discussed in section~\ref{Sec:2D}.

\begin{figure}
\centering
\includegraphics[scale=0.70]{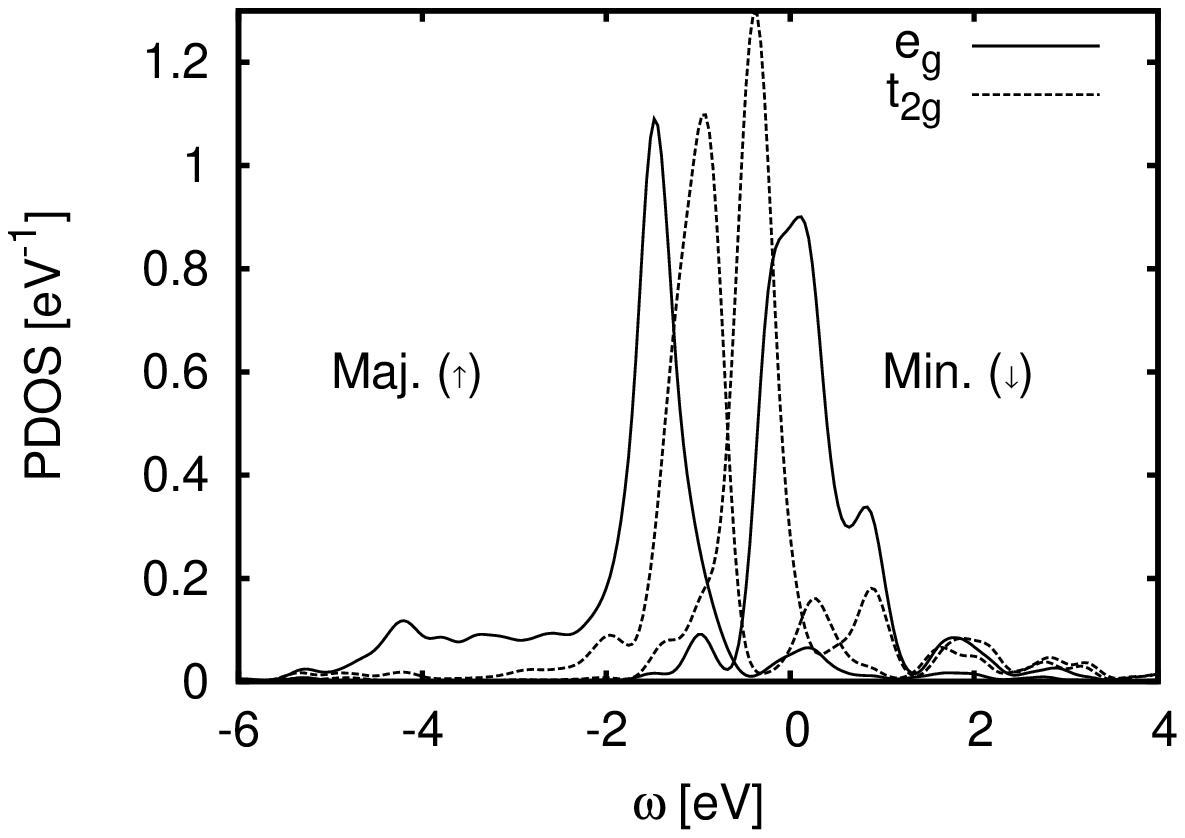}
\caption{\label{Fig_d-wdos_2} Cobalt Wannier-function projected density of
 states in the $4\times4\times4$ cell for both spins. Zero energy coincides with Fermi level.}
\end{figure}

\begin{figure}
\centering
\includegraphics[scale=0.70]{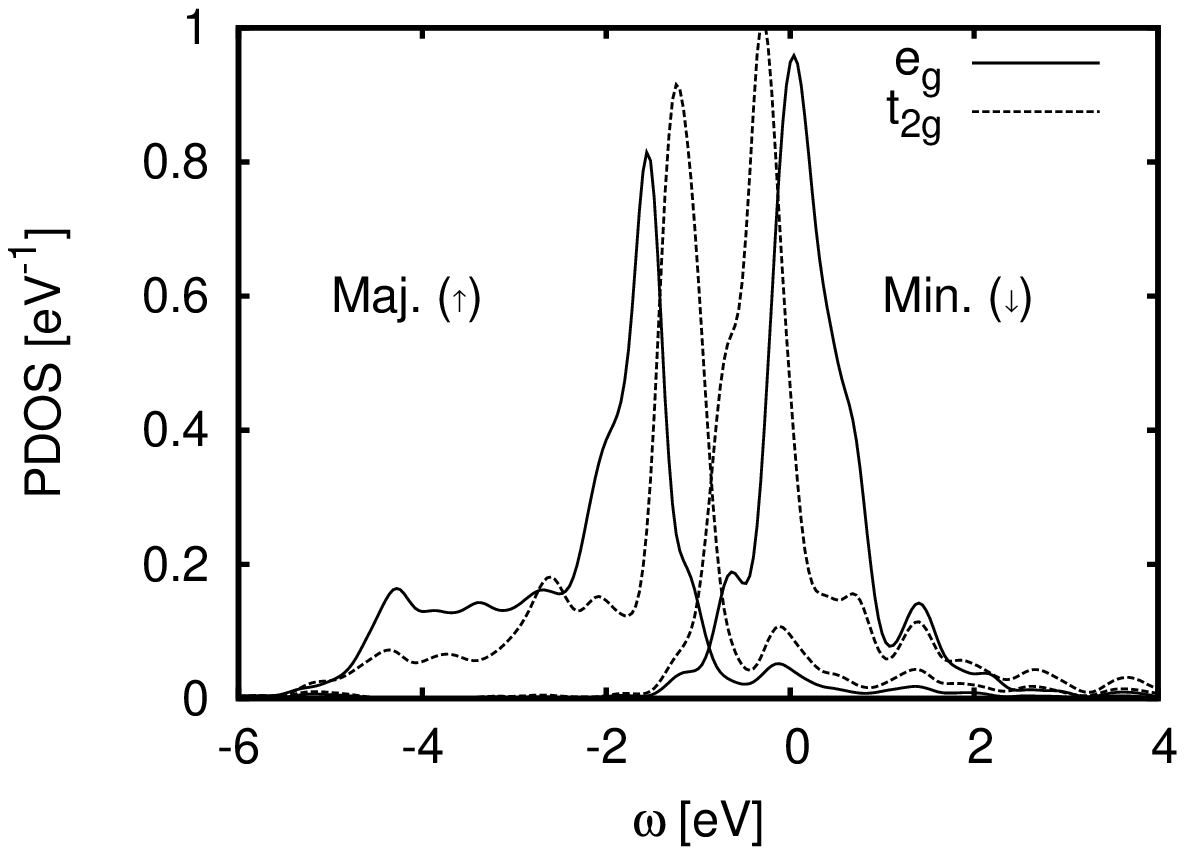}
\caption{\label{Fig_d-wdos_4} Cobalt Wannier-function projected density of
 states in the $4\times4\times4$ cell for both spins. Zero energy coincides with Fermi level.}
\end{figure}

Valuable information is obtained by
analyzing the PDOS in
figures~\ref{Fig_d-wdos_2} and \ref{Fig_d-wdos_4} and
the MLWF occupancies, on-site energies and
real-space spreads as shown in
tables~\ref{Tab_energies_2} and \ref{Tab_energies_4}.

Due to the symmetry of the crystal field~\cite{Goodenough,Stohr},
we find two types of MLWF,  
the $e_g^\uparrow$ one at lower energy, which is twice degenerate according
to the on-site energies (diagonal element of the MLWF Hamiltonian), 
and the one of $t_{2g}^\uparrow$ symmetry, three-fold degenerate. 
We find that the minority spin MLWF's, $e_g^\downarrow, t_{2g}^\downarrow$ 
are actually different in spread from the majority spin
ones, $e_g^\uparrow, t_{2g}^\uparrow$. 
This is typical of unrestricted Hartree-Fock schemes, and
it  reflects the 
fact that both spins correspond to very different energy regions. 
The difference among MLWF's can be seen in the spread, $\Omega$ 
in tables~\ref{Tab_energies_2} and \ref{Tab_energies_4} and in \eref{omega}, 
that measures the extension of the MLWF. We see that $e_g$ MLWF's are more compact
than the $t_{2g}$ ones, and that the majority spin are more compact than the minority
ones. It is interesting to note that the MLWF's are indeed very confined, 
in the present case, the more extended MLWF
is basically zero beyond  2~\AA~from its center. 

The on-site energy is exactly the first moment of the spectral function (\ref{PDOS}),
and hence can be directly correlated to figures~\ref{Fig_d-wdos_2} and ~\ref{Fig_d-wdos_4}.
The occupancy is the integration of the spectral function, over
occupied states, from $-\infty$ to the Fermi energy, \eref{Eq_def_occ}.
Hence, these two quantities help us characterizing the spectral functions.
As shown in figures~\ref{Fig_d-wdos_2} and \ref{Fig_d-wdos_4},
there are four sharp peaks that correspond to the above four types of MLWF.
The presence of tails further from peak centers indicates 
that the MLWF's have an important hybridization with mainly the Cu
host.
These tails shift the on-site energy away from the peak energies.
Most of the weight of the MLWF is, however, concentrated
in a small energy window spanned by the peaks.
Only in the case of the $e_g^\uparrow$ MLWF
does the peak slightly lie outside the {\em inner}
energy window used for the band disentanglement, and develops a tail in the Cu d-band region.

Small differences show up when going from the $2\times2\times2$ cell to the $4\times4\times4$ one.
The main peaks lie at the same energy, and only the first momenta or on-site energies are slightly
different (table~\ref{Tab_energies_4}) because
the spectral function distribution is somewhat narrower for the
smaller cell.
This is due to the smaller content of Co atoms in the large cell, that leads to
smaller Co-Co hoppings.
These comparison permits us to conclude
that the $2\times2\times2$ cell is indeed a good approximation to the dilute
(paramagnetic) regime, in agreement
with the experimental findings by Fan {\em et al.}~\cite{Fan}. 

\begin{figure}
\centering
\includegraphics[scale=0.70]{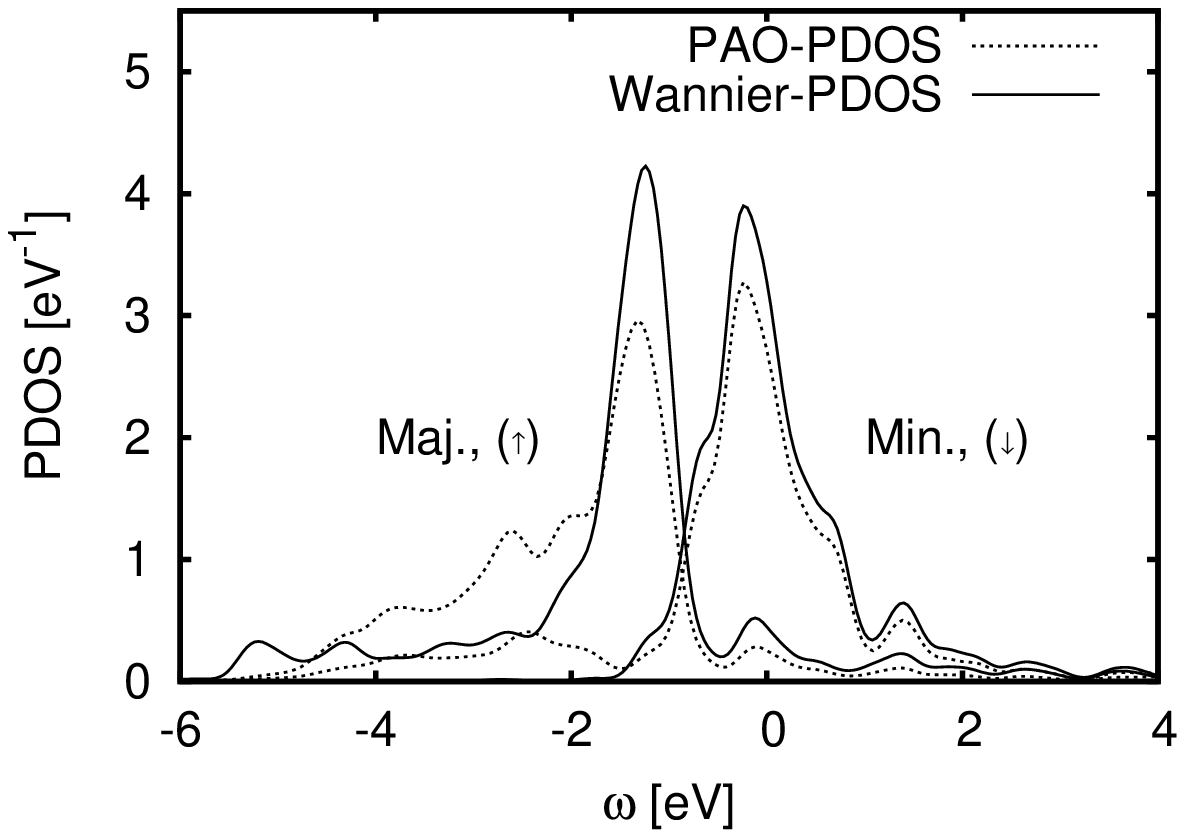}
\caption{\label{Fig_d-wdos-pdos_4}  Comparison of density of states projected on the 
cobalt $e_g$ and $t_{2g}$ Wannier functions and projected density of states onto 
$l=2$ pseudo-atomic orbitals (PAO). The cell $4\times4\times4$ is considered.
Zero energy coincides with Fermi level.}
\end{figure}

The fact that the spectral functions are rather localized in energies permit us to conclude that
they are very good descriptors of the actual Co electronic structure. 
This is further seen
in figure~\ref{Fig_d-wdos-pdos_4}, where the MLWF spectral function is compared with the atomic
basis one ($3d$ Cobalt PAOs). Indeed, the PAO spectral function is more spread in energies, and reveal more
mixing with the Cu-band structure, and, consequently, less Co character.
It is also interesting to notice that the main peaks coincide,
showing that the MLWF contain a lot of physics of the actual electronic structure. 

From the spectral functions we see that the 
different electronic contributions of Co will have 
different physical properties. The $t_{2g}$ electronic structure is basically fully occupied
and slightly contributes to magnetism. However the 
$e_g$ states have a large spin polarization, carrying the leading rh\^ole in the
magnetic properties of the Co impurities.

\subsection{Model Hamiltonian}

The results of the previous section show that the cobalt impurity physics
can be qualitatively understood as
an atomic $d$-orbital weakly hybridized with the Cu substrate.
The hybridization determines the positions and widths of atomic levels,
which in turn fix the impurity occupancy.

There is significant energy splitting regarding
 spin, which results in sizeable
majority and minority spin populations. It is well understood
that the origin of spin polarization 
lies in the intra-atomic Coulomb interaction~\cite{Anderson}.
However, the energy splitting for the $e_g$ levels is
larger than for the $t_{2g}$ ones,
indicating that different $t_{2g}$ and $e_g$
Coulomb matrix elements must be taken into account.

The aforementioned physics can be captured by the multi-orbital
Anderson Hamiltonian\cite{Anderson}
\begin{equation}
\label{Eq_Anderson}
H = H_{at} + \sum_{n\mathbf k}\sum_{m\sigma} 
V_{n\mathbf k,m} c^\dagger_{n\mathbf k\sigma}d_{m\sigma} + H.C. +
\sum_{n\mathbf k\sigma}\epsilon_{n\mathbf k}
c^\dagger_{n\mathbf k\sigma}c_{n\mathbf k\sigma}
\end{equation}
where the atomic part reads
\begin{eqnarray}
H_{at} &=& \sum_{m\sigma}\epsilon_m n_{m\sigma} 
+ 
\half\sum_{m,m'\ \sigma}U_{mm'}n_{m\sigma}n_{m'\bar\sigma},
 \\ 
& & m,m'=1,\dots 5,
\quad
\bar\sigma=-\sigma.
\nonumber
\end{eqnarray}
The substrate electrons (operators $c^\dagger_{n\mathbf k\sigma},c_{n\mathbf k\sigma}$)
with Bloch energies $\epsilon_{n\mathbf k}$ hybridize with impurity given by
$H_{at}$ with on-site energies $\epsilon_m$. Electrons on the impurity site
are subject to Coulomb repulsion with
matrix elements $U_{mm'}$ coupled to orbital occupancies
$n_{m\sigma} = d^\dagger_{m\sigma}d_{m\sigma}$.
Interaction of equal spins is neglected by assuming the same value of direct and
exchange integrals.
The impurity and conduction band are connected by the hybridization term
 with matrix elements $V_{n\mathbf k,m}$.

Construction of the Hamiltonian (\ref{Eq_Anderson}) from first principles using MLWF 
for a cobalt impurity in copper substrate is now discussed.
Orthogonality of Wannier functions allows us to divide the Kohn-Sham Hamiltonian
into the blocks
\begin{equation}
\label{Eq_KS}
\left(\begin{array}{ll}
\mathcal H_{subs} & \mathcal V \\
\mathcal V^\dagger & \mathcal H_{imp}
\end{array}\right)
\end{equation}
where $\mathcal H_{subs}$ is the Hamiltonian in the subspace of substrate
Wannier functions, $\mathcal H_{imp}$ acts on the impurity Wannier functions
(i.~e. the cobalt $e_g$ and $t_{2g}$ functions) while the remaining terms 
$\mathcal V,\mathcal V^\dagger$ are off-diagonal.
In order to be consistent with the mean-field character of the Kohn-Sham DFT
we identify \eref{Eq_KS} with the mean-field solution of the Anderson
Hamiltonian.

Firstly, $H_{at}$ is discussed.
We impose that the on-site energies $\epsilon_m$ obey restrictions of symmetry,
{\em i.e.} have two values which will be denoted by $\epsilon_e$ and $\epsilon_t$ 
for $e_g$ and $t_{2g}$.
In the correlation term, $U_{mm'}$ is a symmetric block matrix of the form
\begin{equation}
\left(
\begin{array}{c|c}
U_{ee} & U_{et} \\\hline
U_{et} & U_{tt}
\end{array}
\right).
\end{equation}

A usual mean-field factorization leads to the simplified Hamiltonian
\begin{equation}
H^{\text{MF}}_{at} = \sum_{m\sigma}\widetilde\epsilon_{m\sigma} n_{m\sigma}
\end{equation}
where the modified atomic energies read
\begin{align}
\widetilde\epsilon_{e\sigma} &= \epsilon_e + U_{ee}\bar n_{e\bar\sigma} + 
 U_{et}\bar n_{t\bar\sigma} \\
\widetilde\epsilon_{t\sigma} &= \epsilon_t + U_{tt}\bar n_{t\bar\sigma} + 
 U_{et}\bar n_{e\bar\sigma}.
\end{align}
The right hand sides are identified with on-site energies of impurity
Wannier functions, for we take $H^{\text{MF}}_{at}=\mathcal H_{imp}$.
The electron number $\bar n_{e\sigma}$ means the total mean occupancy of $e_g$
 symmetry orbitals with spin $\sigma$ and similarly for the $t_{2g}$ states.

To build the Anderson Hamiltonian, the bare energies $\epsilon_m$
and Coulomb integrals $U_{mm'}$ must be determined. Since the occupancies
can be obtained from \eref{Eq_def_occ} we face a linear system of equations.

This system is underdetermined. We solve it by fixing the difference
\begin{equation}
\Delta = \epsilon_t-\epsilon_e = 0.087 eV
\end{equation}
which is the crystal field splitting of atomic levels
that is calculated as the $t_{2g}-e_g$ splitting of bulk copper
$d$-orbitals~\footnotetext{In good agreement with the crystal field splitting in Ag and Au estimated in 
[\cite{Costi}] of about 0.15 eV, and in reference ~\cite{Guo} of less than 0.1 eV.}.
Taking the $4 \times 4 \times 4 $ case, we obtain:
\begin{align}
\epsilon_e &= -4.26 eV, & \epsilon_t &= -4.18 eV,\\
U_{ee} &= 1.63 eV,    & U_{tt} &= 1.19 eV,     \\
U_{et} &= 0.51 eV.    &
\end{align}

Eventhough we are not aware of any estimation of the $U$ for cobalt in bulk copper, values
reported in the literature for bulk Co are about 5 eV~\cite{Nakamura}, and
for cobalt adsorbed on the (111) surface of gold~\cite{Zawadowski} are  2.8 eV. We expect our
values to be smaller than the ones obtainable by the approach of \cite{Nakamura}.
The reason for this is that our approach gives a rule to obtain an
Anderson Hamiltonian from the MLWF Hamiltonian which proceeds
from our DFT calculation and hence has the known problems of the local
and semilocal exchange and correlation functionals in determining $U$ (see for example~\cite{Anisimov91}).
Namely, our  $U$ will correspond to the Hund's rule exchange matrix
element rather than the electrostatic one appearing in LDA+U. 
Nevertheless, our values albeit smaller, are comparable to other $U$
values for Co. Indeed, Antonides and co-workers
measured 1.2 eV for the $U$ of Co as an impurity~\cite{Antonides} 
in excellent agreement
with our calculation. We think that the main advantage of our approach
is that it keeps the complete symmetry and multi-orbital
character of the original DFT calculation.

The construction of substrate and hybridization terms in 
\eref{Eq_Anderson} is straightforward due to their one-particle form.
Firstly, $\mathcal H_{subs}$ is diagonalized, yielding the conduction
band energies $\epsilon_{n\mathbf k}$ and Bloch states. In the next step,
hoppings $\mathcal V$ and $\mathcal V^\dagger$ are transformed accordingly,
leaving us with the matrix elements $V_{n\mathbf k,m}$ between substrate
 Bloch states and impurity Wannier functions.

The essentially complete description of the impurity - substrate mixing
is included in the hybridization function
\begin{equation}
\Gamma_{mm'}(\omega) = \sum_{n\mathbf k}V_{n\mathbf k,m}^{*}
V_{n\mathbf k,m'}\delta(\omega - \epsilon_{n\mathbf k}),
\label{gamma}
\end{equation}
whose diagonal elements $\Gamma_{mm}$ give the spectral intensity of hybridization
of a given impurity Wannier function $m$.
The elements $\Gamma_{mm}(0)$ give (apart from a $2\pi$ factor) the inverse
lifetime of an impurity electron.
Off-diagonal terms provide substrate mediated intra-atomic hybridization
intensity.
The function $\Gamma_{mm'}(\omega)$ is precisely $-\frac{1}{\pi}$ times
the imaginary part of the retarded self-energy due to hybridization.

\begin{figure}
\includegraphics[scale=0.7]{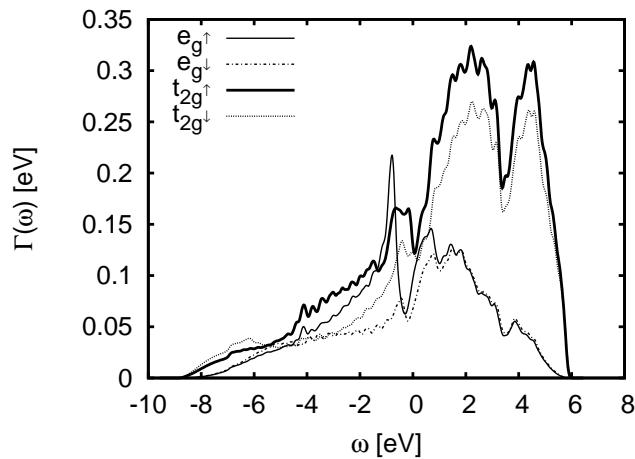}
\caption{\label{Fig_Gamma}Hybridization function, $\Gamma$, calculated for
the cobalt impurity
in the $2\times2\times2$ supercell of copper using a fine Brillouin
zone sampling $40\times40\times40$ and gaussian smearing $0.1$~eV. The matrix
elements corresponding to hybridization of $e_g$ states and $t_{2g}$ states
for both spins are given. Only diagonal contributions (see \eref{gamma}) are shown because the
non-diagonal hybridization functions are orders of magnitude smaller.
Zero coincides with Fermi level.}
\end{figure}

So far we have tacitly omitted spin polarization inherent in the matrices
$\mathcal H_{subs}$, $\mathcal V$ and $\mathcal V^\dagger$, coming from a 
spin polarized Kohn-Sham DFT.
The hybridization function calculated for majority and minority spin
directions is given in figure~\ref{Fig_Gamma}.
Off-diagonal $e_g - t_{2g}$ matrix elements were omitted,
for we found they are of the order of a few meV.
The same holds for elements between different Wannier functions
of the same symmetry.
We see that all intensities have the same order of magnitude.
The $t_{2g}$ functions become much stronger for unoccupied substrate
levels. The hybridization for minoritary spin is weaker than
the majoritary-spin one.
The most pronounced difference is between the $e_g\uparrow$ and 
$e_g\downarrow$ curves. 
The differences between hybridization functions for different spins
partially reflect the polarization of the copper matrix by the magnetic moment
 of cobalt.

\subsection{Surface studies: the case of Cu (111)}

\begin{figure}
\includegraphics{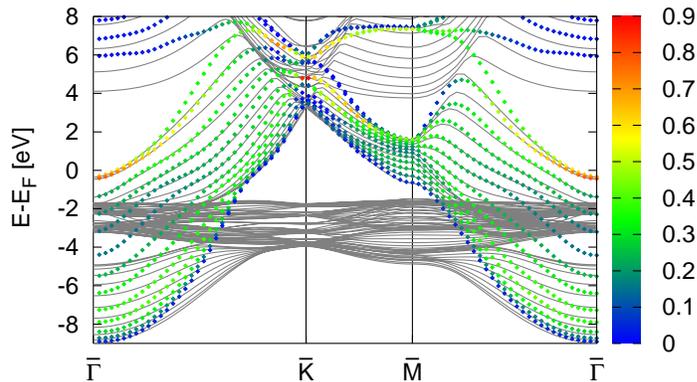}
\caption{\label{Fig_cus-bands} Comparison of \emph{ab-initio} band structure (grey) with the
 Wannier-interpolated (color) of a Cu $(111)$ slab.
 Overlap of the eigenstate with the two outermost surface MLWF is indicated
 by color. The {\em inner} energy  window  spans the
interval from -1.0 eV to 2.0 eV, where the  matching between the interpolated
 and the \emph{ab-initio} bands is very good.}
\end{figure}

Figure~\ref{Fig_cus-bands} shows the comparison of the \emph{ab-initio}
band structure for the Cu (111) and the MLWF interpolated ones.
The disentangling scheme has permitted us to retain the electronic
structure with {\em sp} character about the Fermi energy. As a consequence
neither the {\em d}-bands nor the upper limit of the LL' gap ($\bar{\Gamma}$
for the surface Brillouin zone)
are described by  the MLWF. However, the electronic structure in the
{\em inner} energy window is perfectly
reproduced.

In particular, we underline the excellent description of the Shockley
surface state by the MLWF basis set. The band structure is basically
indistinguishable from the  \emph{ab-initio}  one which in turn
is a very good description of the experimental one~\cite{PRB_Sandra}. 
The minimization procedure leads to two types of differentiated MLWF
sets, one set
describing the bulk electronic structure with a 
spread of $\Omega = 3.3$~\AA$^2$ and a surface MLWF with $\Omega = 5.7$~\AA$^2$
and its original center displaced by 0.21~\AA~into the vacuum region. Hence,
the MWLF try to follow the behavior of the PAO basis~\cite{PRB_Sandra}
where the energy minimization was improved by using two distinct basis sets, one
for the bulk electronic structure and one for the surface with
diffuse orbitals. Indeed, not only does the surface MLWF
reproduce the surface state dispersion (minimum at $E_0=-0.34$ eV, and effective mass $m^* = 0.335$
in electron masses, in
good agreement with $E_0=-0.42$ eV and $m^* = 0.37$ of reference~\cite{PRB_Sandra})
but it also shifts into the vacuum region in
order to account for the surface spilling of charge. 

The interpolated bands can be analyzed in terms of the MLWF character
they have. Figure~\ref{Fig_cus-bands} depicts in a color code the
character of the interpolated bands. In this way we find that
the Shockley surface state is basically purely described by the
surface MLWF near $\bar{\Gamma}$. As we move away from $\bar{\Gamma}$,	
the surface state has some bulk MLWF weight, until it reaches the gap
edge and it becomes a surface resonance with a large bulk MLWF character.

Outside the {\em inner} energy window, the bulk MLWF has more weight. Indeed,
the lowest and highest energy bands are described by the bulk MLWF. This
signals that in this slab geometry, the electronic
structure has mainly a surface character and bulk MLWF are the least
indicated to describe the slab electronic structure.

The MLWF's correspond to s-like orbitals. A consequence of this is
the disentanglement of the electronic structure as seen in figure~\ref{Fig_cus-bands}. There, we see that the MLWF band structure spans from the bottom
of the conduction band and crosses the d-band without mixing with it. Even
outside the inner energy window, the MLWF-bands excellently
reproduce the sp-bands. The description considerably worsens outside the
interpolating {\em inner} energy window. 
 
\section{Summary and conclusions}

We have implemented an interface to the scheme of \cite{CPC_Wannier90} to obtain
a finite set of maximally-localized Wannier functions (MLWF's) from calculations
using the {\sc Siesta} code~\cite{Soler}.

The method yields an efficient band disentanglement in the case of Co impurities
in bulk Cu. We have been able to retain just an $s$-like 
band to represent the Cu electronic structure, while keeping most of the correct
description of the Co electronic structure. As a consequence, we have
analyzed the magnetic properties in the limit of small
concentration of Co impurities, and trace back most of the Co properties
to the ones described by a single MLWF, doubly degenerated with $e_g$ symmetry.

The reduction of the electronic structure to a few important elements is
of great interest to obtain model Hamiltonians from full DFT calculations. We
show that we can map the MLWF Hamiltonian into an Anderson model,
having access to on-site energies, exact hybridization matrix elements and
intra-atomic Coulomb matrix elements that reflect the correct symmetry
of the problem. We show that by using this scheme, we obtain values
of the intra-atomic Coulomb matrix that correspond to the semilocal
exchange and correlation functional, and hence to
the Hund's rule exchange,  that are slightly smaller
than the electrostatic ones obtained in LDA+U schemes~\cite{Anisimov91,Solovyev,Cococcioni}.
These results are encouraging for future studies of strongly correlated systems using MLWF.

Even in the more stringent case of surface electronic structure, the 
description in terms of MLWF give very good results. We have applied
the computational scheme to the Cu (111) surface and inside the
chosen inner energy window the interpolation of the band structure is
excellently reproducing the Shockley surface state with an
accuracy comparable to the complete calculation of reference~\cite{PRB_Sandra},
and permitting us to obtained the sp-band disentangled from the Cu d-bands.
In this way, the electronic structure
problem is limited to the sp-electronic structure
about the Fermi energy.

The surface case shows that not only is the MLWF method a mathematical
trick to disentangle bands and reduce the problem to a smaller, tight-binding
like, basis set. Indeed, one can analyze the obtained electronic structure
in terms of the MLWF and have interesting insight. We have projected
the electronic bands in the two types MLWF of the surface problem: the 
surface  and the bulk MLWF's. In this way, we have verified the
mainly surface character of the Shockley state near the $\bar{\Gamma}$ point,
as well as its subsequent mixing with the bulk MLWF as the surface
state energy increases, until it becomes a surface resonance.

In conclusion, 
we have implemented an interface to the scheme \cite{CPC_Wannier90}  
to obtain a finite set of maximally-localized Wannier functions
from calculations
using the SIESTA code~\cite{Soler}.
In this way, we can obtain DFT-based model Hamiltonians
from \emph{ab-initio}  calculations with a very accurate description of the
electronic structure inside a chosen energy window, that lends itself
to the analysis and exploration of complex systems.

\ack
We thank Prof. P. Gambardella for very interesting discussions and comments.
M.P. thanks Ivo Souza and Jonathan Yates.
Financial support from the Spanish MICINN through 
grant FIS2009-12721-C04-01 is gratefully acknowledged. 
R.K. is supported by the CSIC-JAE predoctoral fellowship.
P.O. acknowledges support from the MICINN CONSOLIDER grant CSD2007-00059.

\bibliographystyle{unsrt}
\bibliography{references}
\end{document}